\documentclass{acmart}
\usepackage{booktabs}
\usepackage{array}
\usepackage{colortbl}
\usepackage{makecell}
\usepackage{graphicx} 
\usepackage{natbib}
\usepackage{titlesec}
\usepackage{float}
\usepackage{verbatim}
\usepackage[colorinlistoftodos]{todonotes}
\newcommand{\etal}{\textit{et al.}}
\usepackage{xspace}
\newcommand{\genai}{Generative AI\xspace}
\usepackage{enumitem}
\usepackage[colorinlistoftodos]{todonotes}
\usepackage{subcaption}
\usepackage{ragged2e}
\usepackage{geometry}
\usepackage{multirow}
\usepackage{caption}
\usepackage{amsmath}

\author{Matin Amoozadeh}
\affiliation{ 
 \institution{University of Houston}
 \city{Houston} 
  \state{TX} 
  \country{United States}
}

\author{Daye Nam}
\affiliation{ 
  \institution{Carnegie Mellon University}
  \city{Pittsburgh} 
  \state{PA} 
   \country{United States}
}
\author{Daniel Prol}
\affiliation{
  \institution{Universidad Internacional de la Rioja}
   \streetaddress{}
  \city{Logroño} 
  \state{} 
   \country{Spain}  
}
\author{Ali Alfageeh}
\affiliation{
  \institution{University of Houston}
  \city{Houston} 
  \state{TX} 
   \country{United States}  
}

\author{James Prather}
\affiliation{
  \institution{Abilene Christian University}
  \city{Abilene} 
  \state{TX} 
   \country{United States}  
}

\author{Michael Hilton}
\affiliation{ 
  \institution{Carnegie Mellon University}
  \city{Pittsburgh} 
  \state{PA} 
  \country{United States}
}

\author{Sruti Srinivasa Ragavan}
\affiliation{ 
  \institution{Indian Institute of Technology}
  \city{} 
  \state{Kanpur} 
  \country{India}
}

\author{Mohammad Amin Alipour}
\affiliation{
  \institution{University of Houston}
  \city{Houston} 
  \state{TX} 
   \country{United States}  
}

\usepackage{fancybox}

\cornersize{.2}
\newcounter{observation}

\title{Student-AI Interaction: A Case Study of CS1 students}

\begin{document}
\renewcommand{\shortauthors}{Matin Amoozadeh et al.}

\begin{abstract}

Generative artificial intelligence tools (\genai), such as ChatGPT, allow users to interact with them in intuitive ways (e.g., conversational) and receive (mostly) good-quality answers. In education, such systems can support students' learning objectives by providing accessible explanations and examples even when students pose vague queries. But, they also encourage undesired help-seeking behaviors, such as by providing solutions to the students' homework. Therefore, it is important to better understand how students approach such tools and the potential issues such approaches might present for the learners.

In this paper, we present a case study for understanding student-AI collaboration to solve programming tasks in the CS1 introductory programming course. To this end, we recruited a gender-balanced majority non-white set of 15 CS1 students at a large public university in the US. We observed them solving programming tasks. We used a mixed-method approach to study their interactions as they tackled Python programming tasks, focusing on when and why they used ChatGPT for problem-solving.  
We analyze and classify the questions submitted by the 15 participants to ChatGPT. Additionally, we analyzed user interaction patterns, their reactions to ChatGPT's responses, and the potential impacts of \genai on their perception of self-efficacy.

Our results suggest that, in about a third of the cases, the student attempted to complete the task by submitting the full description of the tasks to ChatGPT without making any effort on their own. We also observed that few students verified their solutions. We discuss the potential implications of these results.  
\end{abstract}



\maketitle

\section{Introduction}

Many interactive learning environments (ILEs) offer on-demand help to positively influence learning; here, the learner actively seeks information, and the systems provide the same~\cite{aleven2003help}. Although such forms of help-seeking behaviors in ILEs, when effective, are linked to improved learning outcomes, many learners do not use the available help resources effectively\cite{aleven2000limitations,luckin1999ecolab}. This raises concerns that ILEs may not reach their full potential unless we find ways to help students use these support tools better. Furthermore, since seeking help is a crucial skill that affects learning in many situations\cite{nelson1981help,newman2023adaptive,aleven2001investigations}, designing ILEs to encourage effective help-seeking behaviors could greatly enhance their educational value\cite{aleven2001investigations,grasel2000use}. 

\genai can be considered an interactive learning environment (ILE), as
they can provide on-demand help, such as explanations of concepts and working code examples, to students. They play a crucial role as interactive components within broader educational settings, and contribute to interactive learning by providing immediate responses and facilitating learning interactions in real time. \genai tools such as ChatGPT and Copilot have become integral tools for many students, particularly those in introductory programming courses~\cite{prather2023robots}. They can assist students by providing code suggestions, explaining programming concepts, identifying and resolving errors, and generating code documentation~\cite{kazemitabaar2023novices,denny2024cacm}. Their widespread adoption signifies a major shift in how programming is learned and practiced. However, this increasing reliance on \genai tools presents significant challenges. One primary concern is that students might become overly dependent on these tools, potentially hindering their ability to develop fundamental programming skills \cite{prather2024widening}. Additionally, AI-generated solutions might lead to academic dishonesty or reduced problem-solving capabilities among students \cite{becker2023programming}. This dichotomy has sparked a debate among educators: while some argue against the use of \genai in learning due to these risks, others advocate for its judicious use to enhance learning outcomes~\cite{lau2023ban}. 

Despite various studies exploring the impact of \genai tools like GitHub Copilot on productivity~\cite{ziegler2022productivity,kazemitabaar2023studying} and learning outcomes~\cite{tan2024far, margulieux2024self}, there remains a critical gap in understanding how novice programmers interact with these tools without any limitations on using \genai, where they can freely utilize AI-driven assistance and guidance. Understanding the level of trust of students in these tools, which impacts their adoption and learning outcomes~\cite{amoozadeh2024trust,amoozadeh2023towards}, and analyzing their interactions while using \genai for programming tasks is crucial. These insights enable us to identify students' behaviors, actions, decisions, and responses when using \genai tools, guiding the design of  AI-assisted learning interfaces that enhance conceptual understanding and drive improved learning outcomes.
  
To this end, our study investigates the interaction of novice programmers with ChatGPT while solving programming tasks. We conducted a mixed-method study involving 15 CS1 students who used a custom VSCode plugin that integrated ChatGPT directly into their coding environments. This setup allowed us to observe their natural use of \genai assistance without external constraints, and to gain a deeper understanding of how students use \genai in completing programming tasks, namely 1) when, how and why they use them and to what effectiveness, and 2) what strategies they employ to integrate \genai into their problem-solving processes.

Our findings highlight that participants extensively interacted with \genai, yet successfully provided correct answers only in 65\% of the cases; the rest remained unsolved. Some successful interactions involved step-by-step prompting, or hybrid approaches that combine independent programming and \genai support. The overall acceptance of the \genai responses also varied, from full adoption of the \genai's response to using \genai to find answers to the queries, or simply comparing the \genai responses to their own solutions.

\section{Related Work}

\subsection{Exploring the Impact of Generative AI in Programming}
The recent widespread deployment of generative AI programming assistants has motivated much research on the use of these tools. Researchers have conducted empirical studies~\cite{leinonen.2023, Hellas.2023, Sarsa.2022} to evaluate the quality of code and explanations that \genai programming assistants can generate. Others have explored the usefulness of Large Language Model (LLM)-based programming tools through user studies~\cite{ziegler.2022, vaithilingam.2022, barke.2023, kazemitabaar.2023, xu.2022, mozannar.2022, ross.2023,srinivasa2024}. Notably, Vaithilingam \etal ~\cite{vaithilingam.2022} compared the user experience of GitHub Copilot with traditional auto-complete and found that programmers faced more frequent difficulties in completing tasks with Copilot, although there was no significant impact on task completion time. Barke \etal ~\cite{barke.2023} took a step further to understand \emph{how} programmers interact with code-generating models using a grounded theory analysis, identifying two primary modes of interaction: acceleration, where \genai is used to speed up code authoring in small logical units, and exploration, where \genai serves as a planning assistant by suggesting structure or API calls. While these studies have contributed to our understanding of how developers use \genai developer tools, they mainly focused on evaluating the effectiveness of such tools and did not directly address developers' perceptions of them. Some studies, such as those by Liang \etal ~\cite{liang.2023} and Ziegler \etal ~\cite{ziegler.2022}, have also focused on how developers perceive these tools. For instance, Liang \etal ~\cite{liang.2023} conducted a survey involving 410 developers to investigate the usability challenges associated with \genai programming assistants, finding that developers appreciate their autocomplete capabilities, but also reported challenges ranging from the quality of code generation to potentially infringing on intellectual property. Although these studies provide valuable insights on how users \textit{perceive} the \genai tools, the insights are primarily applicable to professional programmers; our study focuses on the use of such tools by CS students, and more studies are needed in this area. 

\subsection{The Role of AI in Education}
Recent studies have started to explore the applications of \genai tools in educational contexts. For example, Denny \etal~\cite{denny2024desirable} have identified desirable characteristics for \genai teaching assistants in programming education. They emphasize the importance of features that provide immediate, engaging support while allowing students to maintain autonomy in their learning process. Liu \etal~\cite{liu2024teaching} investigated the integration of \genai tools in the CS50 course at Harvard University, demonstrating how they can significantly enhance the learning experience for students in introductory programming classes. Their approach involved developing and deploying a suite of \genai-based tools designed to emulate a 1:1 teacher-to-student ratio, thereby providing personalized, real-time support to students. These tools, including the CS50 Duck, a virtual rubber-duck debugging assistant, were well received by students, who reported that the \genai tools made them feel as if they had a personal tutor available at all times. Prather \etal~\cite{prather2024weird} have recently published a series of studies that examine the usability and interaction challenges faced by novice programmers using \genai tools, e.g., GitHub Copilot. Their research highlights the dual nature of such tools: while they can significantly accelerate the coding process and aid in overcoming programming blocks, they also introduce unique cognitive and metacognitive difficulties. Students often found it "weird" how accurately Copilot predicted their needs, which led to a mix of fascination and dependency concerns. The study also identified new interaction patterns such as "shepherding" and "drifting," reflecting the nuanced ways novices attempt to guide \genai or are misled by it. All these prior works together lend motivation to our research, highlighting the potential usefulness of \genai tools, while underscoring the need to balance learning support with independent problem-solving skills and over-reliance on \genai.

\subsection{Interaction Patterns of Novice Programmers Using LLMs}
To develop \genai systems that balance fostering independent thinking with learning support and avoiding overreliance, it is necessary to understand \textit{ how} students use LLMs in practice. In the past, Kazemitabaar \etal ~\cite{kazemitabaar2023novices} conducted a thematic analysis of novice learners using an LLM-based code generator in a self-paced Python course. They identified distinct coding approaches among the learners, noting that the Hybrid approach, which combined manual coding with LLM assistance, produced the best outcomes. However, the study also highlighted signs of over-reliance on LLMs, such as copying LLM output without changes, and positive self-regulation behaviors, like adding code to verify LLM output. In another study, Nguyen \etal~\cite{nguyen2024beginning} conducted a large-scale multi-institutional study that explored the challenges faced by near-novices when interacting with Code LLMs. Their study identified barriers such as difficulties in expressing problem understanding and using appropriate coding terminology. Prather \etal~\cite{prather2024widening} further investigated the impact of generative AI tools on novice programmers' metacognitive awareness and problem-solving strategies, highlighting the potential widening gap between well-prepared and under-prepared students in the era of generative AI. However, there is a lack of studies elucidating when and how students use LLMs, to what effectiveness, and to what effect on their self-efficacy; this is what we cover in the present study.

\subsection{Assessing Student Self-Efficacy in AI-Enhanced Learning}
Self-efficacy, or the belief in one's capabilities to achieve a goal or an outcome, is a crucial factor in students' learning processes. Recent studies have examined how interactions with GenAI tools influence students' self-efficacy in programming. Tankelevitch \etal~\cite{tankelevitch2024metacognitive} discussed the metacognitive demands imposed by GenAI systems, highlighting the need for tools that support students in monitoring and controlling their learning processes. Xue \etal~\cite{xue2024chatgpt} investigated the impact of ChatGPT on introductory programming students, finding that the AI tool can enhance students' self-efficacy by providing immediate feedback and support, thereby reducing the anxiety associated with complex problem-solving tasks. Furthermore, Denny \etal~\cite{denny2024desirable} emphasized the importance of designing AI teaching assistants that not only assist with immediate problem-solving but also help students develop long-term self-efficacy by encouraging independent learning and critical thinking. This study serves to add further evidence on the effect of \genai use on student self-efficacy, albeit using a different approach (namely, pre- and post-task comparisons).

\section{Methodology}

This study focuses on students' use of \genai--covering the whats, whys, whens, and hows of doing so. 
The specific research questions for our study are as follows:

\begin{itemize}

  \item[RQ1:] How frequently do students use \genai while solving programming tasks?
  
  \item[RQ2:] How do students interact with \genai while solving a programming task?
        
    \item[RQ3:] How does the self-efficacy of students change before and after programming with \genai?
\end{itemize}

To answer these questions, we adopted a mixed-methods approach, gathering data from 15 CS1 students using \genai to perform a programming task. As we reason later in this section, the access to \genai in our study was via a plug-in that we designed for the study.

\titlespacing{\subsection}{0pt}{3pt}{1pt}

\subsection{Participants}
We recruited students from a CS1 course in a large US public university. We sent an email to the course email list, via the instructor (also an author of this paper), inviting volunteers to participate in a research study involving programming tasks with AI. Each participant was offered \textit{five} bonus points in their course for their participation. We considered offering bonus points so as to not coerce students into participating. Participants could simply fill in an available time slot for the study if they were willing to participate. The course had 110 students, of which 15 students volunteered to participate. 

Table \ref{tab:participant-demographics} summarizes the demographics of the participants. Our study was gender balanced (8 females, 7 males). Two of our participants were first-generation students (i.e., no parent or guardian possessed a four-year college degree); the rest were continuing-generation students and had at least one parent or guardian with a four-year college degree. Most of our participants (13 out of 15) were freshmen; the other two were junior and post-baccalaureate respectively. The participants reported diverse racial/ethnic backgrounds, and everyone used English as their first or second language fluently.


\begin{table}
    \centering
    \begin{tabular}{lcccccc}
        \toprule
        Participant & Age & Gender & Education-level & Racial/Ethnic & College-generation\\
        \midrule
        P1 & 19 & Female & Freshman & African American & Continuing-generation &\\
        P2 & 18 & Female & Freshman & African American & Continuing-generation &\\
        P3 & 19 & Female & Freshman & African American & Continuing-generation &\\
        P4 & 19 & Male & Freshman & Asian & Continuing-generation &\\
        P5 & 18 & Female & Freshman & Asian & Continuing-generation &\\
        P6  & 19 & Female & Freshman & Asian & First-generation &\\
        P7 & 19 & Male & Freshman & African American & Continuing-generation &\\
        P8 & 18 & Male & Freshman & Latino & First-generation &\\
        P9 & 22 & Female & Postbacc &African American & Continuing-generation &\\
        P10 & 18 & Male & Freshman & Asian & Continuing-generation &\\
        P11 & 19 & Female & Freshman & White & Continuing-generation &\\
        P12 & 19 & Male & Freshman & African American & Continuing-generation &\\
        P13 & 18 & Male & Freshman & African American & Continuing-generation &\\
        P14 & 20 & Male & Junior & Latino & Continuing-generation &\\
        P15 & 19 & Female & Freshman & African American & Continuing-generation &\\
        \bottomrule
    \end{tabular}
    \caption{Demographic Information of Participants}
    \label{tab:participant-demographics}
\end{table}

\subsection{Study Procedure}

The study was conducted in an office within the computer science department during the spring semester of 2024, with one participant per study session. The moderator conducting the study was not part of the CS1 course in any capacity (instructor, tutor, or TA). 

Each study session began with a study debriefing, and the participant signing the informed consent form. The participant then filled out a pre-study questionnaire, which contained questions about: 1) demographics, 2) programming background, and 3) self-efficacy questionnaire(from~\cite{pintrich1991manual}).

Then, the participants were introduced to the study environment. We chose the Python programming language and VSCode environment for the study, since it was already used in the course. Participants were also introduced to the plug-in, and were briefed on how to ensert prompts into it, should they choose to use \genai for their tasks. We also emphasized to participants that other \genai or aids (e.g., web searches) were not allowed during this time. We also clarified that their performance would not be judged by others, or affect their course grades, to ensure that the participation was non-coercive. 

Participants were introduced to the programming tasks on a sheet of paper, and were given one hour to complete the tasks. 
On completion of the tasks, the participants completed a post-study survey containing the same self-efficacy questionnaire administered as part of the pre-study survey. The study sessions lasted between 30 and 70 minutes each.

\subsection{Programming tasks}
Participants were asked to complete three programming tasks, each of which is tailored to the complexity appropriate for a CS1 level and covered topics taught by the instructor throughout the semester preceding the study.

Participants were provided with input and output examples for each task. 
Collectively, these three tasks addressed core programming concepts such as conditionals, loops, functions, input/output operations, and simple algorithm design, providing participants with a comprehensive understanding of essential programming principles and problem-solving approaches.

As mentioned earlier, participants had a choice of using \genai or not to complete these tasks. The use of \genai had to be via the custom plug-in we developed, and the use of any other resource (e.g., web search) was disallowed. 

\subsection{\genai plugin}
The goal of this study is to understand when, how, and why CS1 students interact with \genai tools and how effective these interactions are. Therefore, we decided to allow only the use of \genai for students' help seeking to complete their tasks and disallowed all other sources.

In deciding what \genai tool to use for the study, we had several choices: OpenAI ChatGPT, Google Gemini, and Github Copilot, to name a few. We chose to use OpenAI ChatGPT for two reasons: 1) it provided general help beyond programming, and 2) from a prior study, it seemed to be the most familiar \genai among students \cite{amoozadeh2024trust}.

However, in a recent study, Tankelevitch \etal~\cite{tankelevitch2024metacognitive} highlighted the metacognitive demands placed on users when interacting with \genai systems (e.g., the need to remember sub-tasks and which one they are on), and suggested the need to minimize switching between the task and the \genai help-seeking contexts. Therefore, we used a GPT-3.5 plug-in within the VSCode\footnote{https://code.visualstudio.com/}~\cite{Daye:ChatGPT:Plugin} programming environment, to reduce the cognitive load associated with switching between different environments, thereby allowing participants to remain within a single, cohesive coding environment. 

\begin{figure}
\centering
\includegraphics[width=0.9\textwidth]{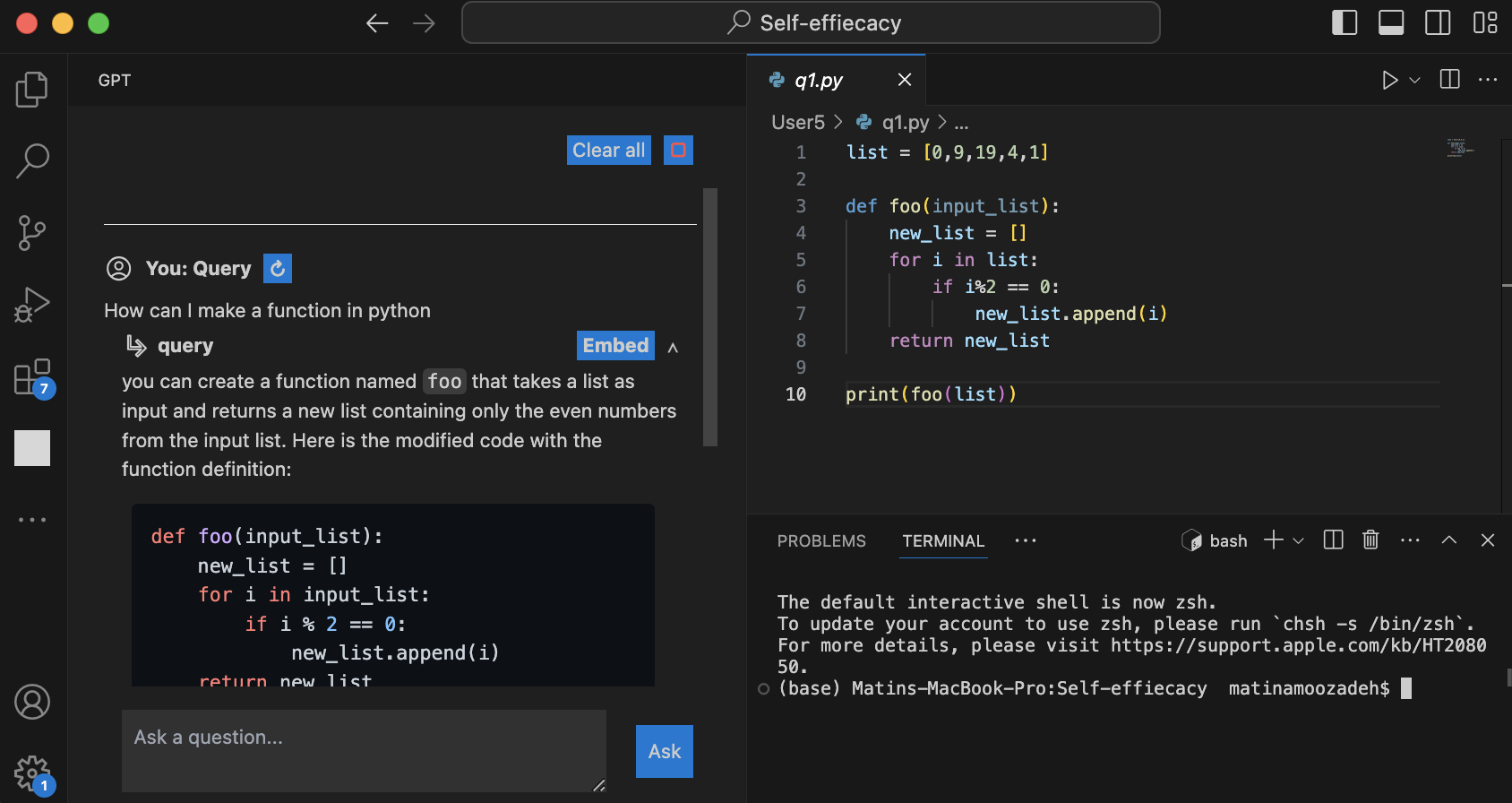}
\caption{Interface of the VS Code plug-in used in the study: ChatGPT Response and Participant Code}
\label{fig:VS.png}
\end{figure}

Figure~\ref{fig:VS.png} shows the VSCode plugin ~\cite{Daye:ChatGPT:Plugin} we used for this study. As the figure shows, the plugin had a simple interface comprising of a prompt-writing box (See "Ask a question..." in the image), allowing users to access ChatGPT within the VSCode IDE. This prompt box allows students to freely input prompts without any restrictions on the number or type of questions, or prompt lengths. The \genai responses to the prompts are also provided in the same pane, as the figure shows. The fact that users can access \genai and see its responses together with their code in the same window allows for a focused environment where users do not need to switch context to another window, and are less tempted to use other AIs. 

\subsection{Data gathering}
Our study employed a multi-faceted approach to data collection, designed to capture a comprehensive view of participants' interactions with \genai (here, ChatGPT) during programming tasks. We utilized a custom VSCode plugin to log all text editor interactions (including any text selections) and ChatGPT prompts, and responses. The audio and video recordings of the participants, along with recordings of the participants' screens throughout the session provided qualitative data on problem-solving behaviors and reactions to AI assistance. Pre-study and post-study surveys collected demographic information and self-efficacy data using Likert scales. In addition, we collected detailed logs of ChatGPT conversations, including the full text of prompts and responses. This diverse dataset allowed us to analyze the frequency and timing of \genai use (RQ1), students' activities and their patterns (RQ2), their \genai use strategies (RQ3), and the effects of \genai use on students' self-efficacy (RQ4). By combining automated logging, video analysis, and self-reported measures, we ensured a foundation for both quantitative and qualitative analysis of novice programmers' engagement with AI-assisted coding.

\subsection{Data analysis}
To answer our research questions, we conducted both qualitative and quantitative analyses of the data. 

\subsubsection{Qualitative.} Two authors independently coded the video data for three programming tasks, each varying in difficulty levels and topics, utilizing the initial codebook. Subsequently, they engaged in discussions to review the initial coding outcomes, resolve any conflicts, and enhance the codebook further. With the identification of each new code, we reviewed all prior comments for possible adjustments.

The codebook provides a clear breakdown of how participants interacted with \genai prompts throughout various stages of programming tasks. 
It categorizes prompts based on their types and how participants engaged with them, whether standalone or as follow-up interactions. Additionally, it outlines how participants responded to errors during programming and their preferences for resolving them, including using \genai responses or ignoring errors altogether. Moreover, the book details the extent to which participants accepted AI-generated solutions, from copying entire responses to incorporating only certain parts or using them for guidance. 
In essence, the codebook served as a guide for understanding participant-AI interactions during programming activities in educational settings. 
The codebook includes such thematic coding of responses allowing us to draw meaningful insights about students' perspectives and shed light on the complex range of attitudes and emotions surrounding \genai systems in the context of programming.

\subsubsection{Quantitative.}
We conducted two surveys, one administered before the programming session and another after, to assess participants' level of self-efficacy. Participants rated their self-efficacy levels using 5-level Likert scales, indicating their agreement with statements related to their programming abilities. These Likert scores provided quantitative data to measure changes in participants' self-efficacy before and after engaging in the programming tasks.
  
\section{Results}
Fifteen CS1 students participated in our study, and we asked each of them to do their best to complete three programming tasks. Table~\ref{tab:usage-summary} summarizes the tasks completed and the correctness of solutions for each participant. Notably, most participants (P3, P15) had atleast one correct solution,  but only 6 out of 15 participants (P3, P5, P6, P9, P12, P14) solved the three programming tasks successfully. Overall, we had 40 completed (correct + incorrect) solutions submitted by participants. We then analyzed the logs from the \genai. 

\subsection{RQ1:  How frequently do students use \genai while programming?}

We used the logs from the plugin to analyze the frequency of \genai use. Of the 40 completed participant submissions, 29 solutions were created with assistance from the \genai plug-in provided, and 11 without \genai. requests, as a measure of the frequency of participants' \genai use during each task. Table \ref{tab:usage-summary} displays this frequency of participants' plugin-aided \genai use, illustrating the different levels of dependence on \genai assistance among participants.

\begin{table}
    \centering
    \begin{tabular}{lcccccc}
        \toprule
        Participant & Tasks & Completed & Correct & Using \genai & \#Times \genai \\
        &  & Q1,Q2,Q3 & Q1,Q2,Q3 & Q1,Q2,Q3 &  \\
        \midrule
        P1  & Condition, String, Function & Y, Y, N & W, W, N  & N, Y, N & 0, 1, 0 \\
        P2  & Condition, String, Function & Y, Y, Y & W, W, C & Y, Y, Y & 9, 3, 1 \\
        P3  & Condition, String, Function & Y, Y, Y & C, C, C & N, N, Y & 0, 0, 2 \\
        P4  & Condition, String, Function & Y, Y, N & C, W, N & N, N, N & 0, 0, 0 \\
        P5  & Condition, String, Function & Y, Y, Y & C, C, C & N, Y, Y & 0, 1, 10 \\
        P6  & String, List, Array & Y, Y, Y & C, C, C & N, Y, N & 0, 5, 0 \\
        P7  & String, List, Array & Y, Y, Y & C, C, W & Y, Y, Y & 2, 1, 4 \\
        P8  & String, List, Array & Y, Y, Y & C, C, W & N, N, Y & 0, 0, 3 \\
        P9  & String, List, Array & Y, Y, Y & C, C, C  & Y, Y, Y & 1, 1, 1 \\
        P10 & String, List, Array & Y, Y, Y & C, C, W & Y, Y, Y & 1, 2, 1 \\
        P11 & String, List, Array & Y, Y, Y & W, C, C & N, Y, Y & 0, 1, 1 \\
        P12 & String, List, Array & Y, Y, N & C, C, N & Y, Y, N & 3, 1, 0 \\
        P13 & String, List, Array & Y, N, N & W, N, N & Y, N, N & 3, 0, 0 \\
        P14 & String, List, Array & Y, Y, Y & C, C, C & Y, Y, Y & 1, 1, 1 \\
        P15 & String, List, Array & Y, Y, Y & W, W, W & Y, Y, Y & 1, 1, 1 \\
        \bottomrule
    \end{tabular}
    \vspace{5pt}
    \caption{Usage and Problem Solving by Participants: This table details participants' problem-solving activities. ``Participant'' identifies participants, ``Tasks'' shows the main concepts in the tasks attempted, and ``Completed'' indicates if the task has been completed by the participant  (`Y' for yes, `N' for no). ``Correct'' shows if the submission was correct (`C'), wrong (`W'), or not attempted (`N'). "Using \genai" indicates \genai use (`Y' for yes, `N' for no), and ``Times \genai'' shows the number of times \genai used.}
    \label{tab:usage-summary}
\end{table}

However, these use frequencies did not always translate into task completion successes. Overall, among the 29 \genai assisted solutions created by participants, only 19 solutions ($65\%$) were correct and the rest (35\%) were incorrect. Even among two participants (P2, P5) used \genai extensively for a task, only one instance (P5, Q3) was successful; P2 did not manage to complete Q1 correctly, even after 9 \genai queries. Moreover, some participants, such as P3 and P6, achieved correct answers ('C') for one or all tasks even without relying on \genai, indicating that some participants may not need \genai to perform well. 

Thus, participants' frequency of \genai use alone did not lead to successes; instead, participants' strategies in using \genai and their inherent problem-solving abilities both played crucial roles in their success.

\subsection{RQ2: How do students interact with \genai while programming?}
To understand how and why some participants were able to use \genai to correctly complete tasks, and others were unsuccessful in doing so, we delved into various qualitative aspects of participants \genai use--when and how participants interacted with \genai, and how these translated into task completion successes. 

\subsubsection{When do students use \genai?} We observed that participants typically engaged with generative AI at three different stages during their task:

\begin{itemize}
\item \textit{Early in the task.} 
In our study, 8 participants, in a total of 18 tasks, turned to ChatGPT right away, at the beginning of the task, without making any initial attempts on their own; see instances of "Early" in Table~\ref{tab:ai-usage-approach} under "When". Of them, four participants (P9, P10, P14, P15) used \genai right from the beginning for all three programming tasks. Typically, during these instances, participants typed in the description of the task as is, and relied on \genai to complete the task for them.

\item \textit{Middle of the task.}
In 9 out of 45 total tasks, participants (7 out of 15) sought the assistance of \genai in the middle of their programming; these instances are labeled "Middle" in Table~\ref{tab:ai-usage-approach}, under "When". Initially coding manually, they resorted to using \genai when they encountered errors, or to understand programming concepts. 

\item \textit{After completing the task.}
Interestingly, two participants (P3 and P5) employed \genai even after they had already arrived at the correct answer on their own. See instances of "After-Solving" in Table \ref{tab:ai-usage-approach}; Section "When". Here, P3 used \genai to validate she understood the problem description correctly by inquiring about the distinction between the terms (i, N) mentioned in the problem. P5 reached the correct solution on their own first, but still sought \genai's solution for the entire question, apparently to verify the correctness and/or effectiveness of their solution by comparing their solution with the \genai's responses.

\end{itemize}

\begin{table}
    \centering 
    \begin{tabular}{l*{9}{c}} 
        \toprule
        \textbf{User} & \multicolumn{3}{c}{\textbf{When (\genai Used)}} & \multicolumn{3}{c}{\textbf{Approach}} & \multicolumn{3}{c}{\textbf{Usage Pattern}} \\
        \cmidrule(lr){2-4} \cmidrule(lr){5-7} \cmidrule(lr){8-10} 
         & \textbf{Q1} & \textbf{Q2} & \textbf{Q3} & \textbf{Q1} & \textbf{Q2} & \textbf{Q3} & \textbf{Q1} & \textbf{Q2} & \textbf{Q3} \\ 
        \midrule
        P1  & - & Middle & - & - & Hybrid/partial & - & I & I & - \\ 
        P2  & Early & Middle & Early & step-by-step & Hybrid/partial & Full description & I & I & I \\ 
        P3  & - & - & After-Solving & - & - & step-by-step & I & I & I \\ 
        P4  & - & - & - & - & - & - & I & I & - \\ 
        P5  & - & After-Solving & Early & - & Full description & step-by-step & I & L & I \\ 
        P6  & - & Middle & - & Themselves & Hybrid/partial & - & L & I & I \\ 
        P7  & Middle & Middle & Middle & Hybrid/partial & Hybrid/partial & Hybrid/partial & I & L & I \\ 
        P8  & - & - & Middle & - & - & Hybrid/partial & I & I & I \\
        P9  & Early & Early & Early & Full description & Full description & Full description & L & L & L \\ 
        P10 & Early & Early & Early & Full description & Full description & Full description & L & I & L \\ 
        P11 & - & Early & Early & - & Full description & Full description & L & L & L \\
        P12 & Early & Middle & - & Full description & Hybrid/partial & - & I & I & - \\ 
        P13 & Middle & - & - & step-by-step & - & - & I & I & - \\ 
        P14 & Early & Early & Early & Full description & Full description & Full description & L & L & L \\ 
        P15 & Early & Early & Early & Full description & Full description & Full description & L & L & L \\ 
        \bottomrule
    \end{tabular}
    \caption{Summary of \genai Usage and Approaches by Participants: The table summarizes the \genai usage and approaches taken by users for different programming questions (Q1, Q2, Q3). The first set of columns under "When (\genai Used)" indicates the timing of \genai usage classified as "Early", "Middle", "After-Solving", or "-" if not used. The second set of columns under "Approach" details the approach taken by participants, such as "Hybrid/partial", "Step by step", "Full description", and "-" if not used. The third column shows nature of participants' activities, with some repeating activities multiple times as iterative "I" and others performing them only once as linear shown as "L".}
    \label{tab:ai-usage-approach}
\end{table}

\subsubsection{What are common \genai usage patterns among students?}
We conducted an activity analysis of the students' programming tasks, coding the activities they engaged in. We identified six common activities users performed when using \genai, namely, reading, thinking, writing code, modifying code, prompting, and debugging. Table \ref{tab:actions} defines these activities. Participants engaged in these activities in varying levels, with some engaging in the same activity on multiple instances during the same task, and others performing them only once per task.

\begin{figure*}
    \centering
    \begin{subfigure}{0.48\textwidth}
        \raggedright
        \includegraphics[width=\textwidth]{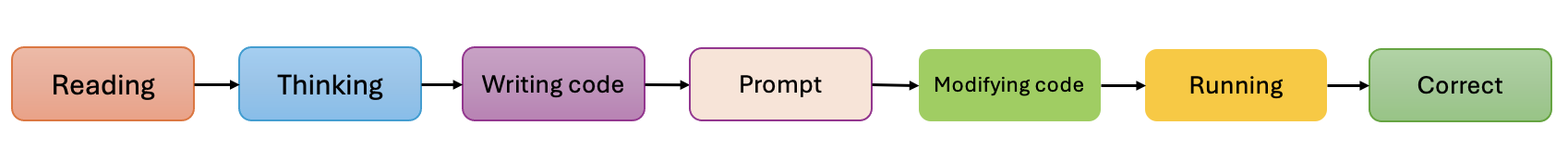}
        \caption{Linear activities of P14 for all questions}
        \label{fig:P14}
    \end{subfigure}
    \hfill
    \begin{subfigure}{0.48\textwidth}
        \raggedright
        \includegraphics[width=\textwidth]{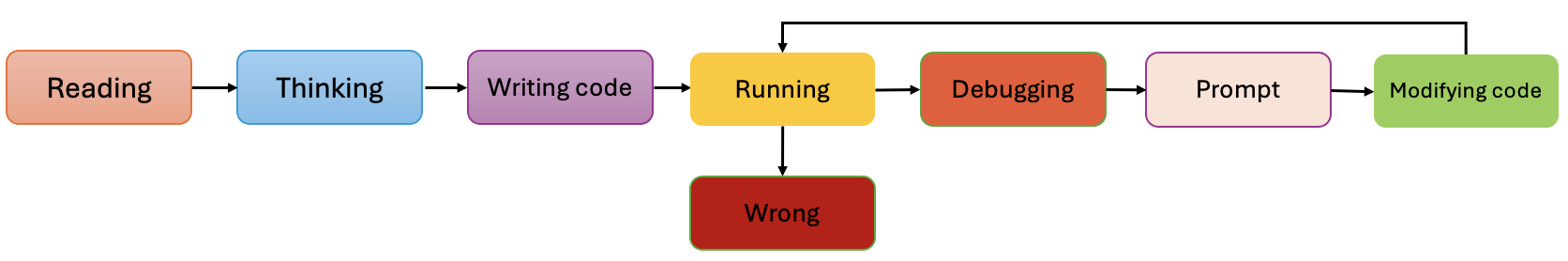}
        \caption{Iterative activities of P7 for Question 3}
        \label{fig:p7}
    \end{subfigure}
    \caption{Examples of activities}
    \label{fig:combined}
\end{figure*}

An analysis of the activity sequences for each task revealed two large patterns, as illustrated in \ref{fig:combined}; the figure shows the activity sequences\footnote{Notice that in the figure "correct" and "wrong" are not activities; we also did not count "running code" as an explicit activity because it was a part of task completion or verification.} of two tasks by P14 and P7 respectively. 
Whereas P14 followed a straight path across all three programming questions, performing each activity once, P7 in question 4 repeated certain activities such as running, debugging, prompting, modifying, and running again. Overall, the 15 participants in our study completed a total of 40 tasks (Table~\ref{tab:usage-summary}, column Completed); of them, we observed 17 instances of straight or \textit{linear} activity sequences and 23 instances of repetitive activities. These are indicated by ``L'' and ``I'' in Table \ref{tab:ai-usage-approach}, under "Usage Pattern".

\begin{table}
    \centering
    
    \begin{tabular}{@{}ll@{}}
        \toprule
        \textbf{Activity} & \textbf{Description} \\
        \midrule
        Reading & Reading the question before starting to write code \\
        Thinking & Thinking about solution, or \genai prompt and response \\
        Writing Code & Writing code by themselves or copying \genai response \\
        Modifying Code & Editing existing code \\
        Prompting & Asking questions from \genai \\
        Verification and Debugging & Running code and finding errors \\
        \bottomrule
    \end{tabular}
    \caption{Description of six common activities users performed when using \genai user activities.}
    \label{tab:actions}
\end{table}

The pattern of a student returning to the same activity is an indicator of an iterative process (e.g., several iterations of re-running code, or prompting \genai more than once for the same task). In contrast, a linear pattern is an indicator of one-shot task completion, assisted with or without \genai. The fact that over 40\% instances were linear is suggestive of one of two possibilities. The first is that of high task performance abilities, as marked by minimal help-seeking, and lack of edit-verify loops. The second possibility is that of over reliance on \genai, wherein participants made a single prompt to \genai the result of which resulted in a direct solution to the problem, following which participants did not have to enter an edit-verify loop. To assess which of these possibilities caused the observed usage patterns, we drilled down into the nature of participants' interactions with \genai.

\subsubsection{What interaction strategies do students adopt with \genai?}
There were three key aspects to participants interactions with \genai: the level of task decomposition, the kind of information they needed, and how participants exploited the \genai response.

\paragraph{Levels of task decomposition}
As Table \ref{tab:ai-usage-approach} shows, participants engaged in three prompting strategies, based on the level of task decomposition: \textit{full description}, \textit{step-by-step}, and \textit{partial}: 


\begin{itemize}
    \item In the \textit{full description} strategy, the participant copied the full description of the task into the plugin, thereby offloading task completion entirely to \genai. In total, there were 17 instances across 9 participants where participants, early in the task (Table~\ref{tab:ai-usage-approach}), provided the entire problem description to \genai to be solved by the latter. Of them, 5 participants adopted this approach for only one or two tasks, and relied on step-by-step or hybrid /partial approach for others. However, 4 participants (P9, P10, P11, P14, P15) entirely relied on this strategy for all tasks, taking to \genai very early in the task. Note, however, that this opportunistic approach to offload the task completion to \genai did not translate to complete and correct solutions. Among the participants who relied on full descriptions, P10 (for Q3) and P15 (for Q1, Q2, Q3) were unable to find the correct answers. 

    \item In the \textit{step-by-step} interaction strategy, participants broke the question down into structured sub-goals and then used the \genai to solve the problem step by step--resulting in an iterative approach. Four participants (P2, P3, P5, P13) adopted this problem-solving strategy. For example, participant P5 used \genai for the first time to inquire, 'How to round numbers in Python?' and then followed up with a second query, ``How to round numbers in Python only from the tenths place?''. Similarly, P13 used \genai three times for their first question. The first inquiry was, ``In Python, can you turn an integer into a list?'' The second query was about arrays: ``What is an array?'' Lastly, the user asked: ``Given an integer n, produce the array: 0123…n, 1234…0, n012…n-1?.'' Indeed, there is diversity in the nature of prompts, as we describe later.
    
    \item The remaining participants used a \textit{hybrid} or partial strategy where participants tackled some parts of the task independently while seeking assistance from \genai for some other parts. Notably, participants attempted to write programs independently, but turned to \genai for debugging to resolve errors or others. We identified n=8 instances that we categorized as Hybrid or Partial. For example, when stuck with a bug, participant P2 asked \genai, ``Fix my code to work.'' User P6 encountered an error and copied the message: \texttt{"TypeError: 'int' object is not subscriptable"}. We observed only one instance of a participant successfully completing a task on their own (Table~\ref{tab:ai-usage-approach}, Q1, P6 labeled "Themselves").

\end{itemize}

\paragraph{Information needs.}
We analyzed the prompts that users submitted to ChatGPT, via the logger.  In all, 15 participants wrote 60 prompts, and we categorized them based on the information need the prompt aimed to serve.
\begin{itemize}
\item \textit{Entire Solution.} As described earlier, several participants simply typed in the question descriptions, which we considered as their entire prompt. These accounted for a third of all prompts (20 out of 60). Often, these prompts were issued early on during the task, and participants obtained the entire solution to the task, via this single prompt; 8 out of 15 participants engaged in this behavior.
\item \textit{Coding Concepts.}
Another popular prompting strategy, namely seeking to understanding programming concepts, was equally common among participants. About a half of participants (7 out of 15) accounting sought such information, accounting for ~30\% (19 out of 60) of all prompts. These instances of conceptual understanding, arose in largely two cases, namely when the participant had a logical understanding of what to do, but needed help executing them in Python (e.g., P2: ``How to append to a list?'', P5:``How to round numbers in Python?''), to understand whether the language allowed something (e.g., P7: ``In my program, I am trying to move the negative numbers to the front. Could I initialize an empty list first?''), or to understand jargon (e.g., P13: ``What is an array?').

\item \textit{Program Logic. }In about 15\% of the time (10 out of 60 prompts), participants utilized \genai because they found it challenging to determine the sequence of steps or instructions needed to accomplish a specific task or goal within the program. In other words, they used \genai for planning help. For instance, P2 asked about "a function to return the first n values of the triangular number sequence starting from 1.", expecting to receive the steps to accomplish the task. P12 needed help trying to reverse a list. P7 inquired, "I am trying to produce the array 0, 1, 2, 3 all the way to n, meaning it could also be n, 0, 1, 2. Is my approach correct so far?".

\item \textit{Debugging.} Finally, some participants turned to \genai for debugging and their questions typically revolved around the question: ``How can I solve this error?''. In total, 11 out of 60 prompts utilized \genai for error resolution. Often, users would simply copy the error message from the console and ask ChatGPT for help. P6 copied and pasted the code, asking, ``What is wrong with this code?'' P10 asked, ``Why does the code not work?'' P6 specifically inquired about the error received, in some ways treating it like a search engine: \texttt{``TypeError: ‘int’ object is not subscriptable.''}

\end{itemize}


\paragraph{Exploiting \genai responses}

We identified various user actions upon receiving responses from \genai, which we categorize as acceptance categories. These actions reflect users' reactions to \genai answers, and this collaboration can result in either successes or failures in the completion of tasks. These categories are based on the prompts and their reactions to the \genai's answers.

\begin{itemize}
\item \textit{Entire response.} Participants often accepted the \genai's response in its entirety. We had $n=18$ out of 60 prompts we found that users accepted \genai answers, without any evidence of explicitly evaluating the responses. For example, P10 formulated a prompt for Question 3 ``Given an integer n, produce the array'' and accepted the \genai's solution without any modification. Similarly, P7, when addressing the same question, queried ``I am trying to produce the array 0,1,2,3 all the way to n, meaning it could also be n,0,1,2 is my correct so far?'' fully embraced the \genai's response.
 
\item \textit{Selective use. }Another common use of \genai responses was that participants tried to understand the \genai response, and then translate that understanding to implementing their own solution. There were $n=23$ out of 60 prompts where users sought ideas from \genai responses to write their own code. For instance, P5 asked: ``How to round numbers in Python only from the tenths place'' and the \genai response was ``To round numbers in Python only from the tenths place, you can use the round() function'', followed by an example of how \texttt{round()} function works. The user then completed the task using their own code.

\item \textit{Reject and Retry.} Using recorded videos of facial expressions and eye movements, we also identified cases in which participants spent an extended amount of time reading and (most likely) contemplating \genai's responses. We observed six participants visibly contemplating the \genai's response to 19 (out of 60) prompts. These participants read the responses but did not exploit it, by way of either copying the code from the response, or simply write it in their own way. Instead, they went on to write another prompt, which may indicate their rejection of the \genai solution The reasons for this varied from \genai responses not meeting their information needs, to a rejection of the implementation choices made in the responses. P12's initial query was ``I need help trying to reverse a list.'' After reading the \genai response, the user attempted to write another prompt, ``Help me turn a sequence of numbers into a list.'' Once again, the user read the \genai's response. Subsequently, P12 revised his prompt and wrote: ``Without using a built-in function, help me turn a sequence of numbers into a list.'' After this prompt, he began to write code in their own way.

\end{itemize}

\subsection{RQ3: How does student self-efficacy change before and after programming with \genai?}
With such diversity in \genai usage among participants--in terms of frequencies, usage patterns and strategies, and success rates-- we went beyond simple task completion to metacognition, specifically to evaluate the impact of \genai tools on students' self-efficacy.  

For this, we analyzed the self-efficacy questionnaires we administered as part of the pre-study and post-study surveys. Unfortunately, for the first five participants (P1 to P5), we did not capture the pre-study self-efficacy data; thus, we had data from 10 participants. 

\begin{figure}[ht]
    \centering
    \begin{subfigure}[b]{0.45\textwidth}
        \centering
        \includegraphics[width=\textwidth]{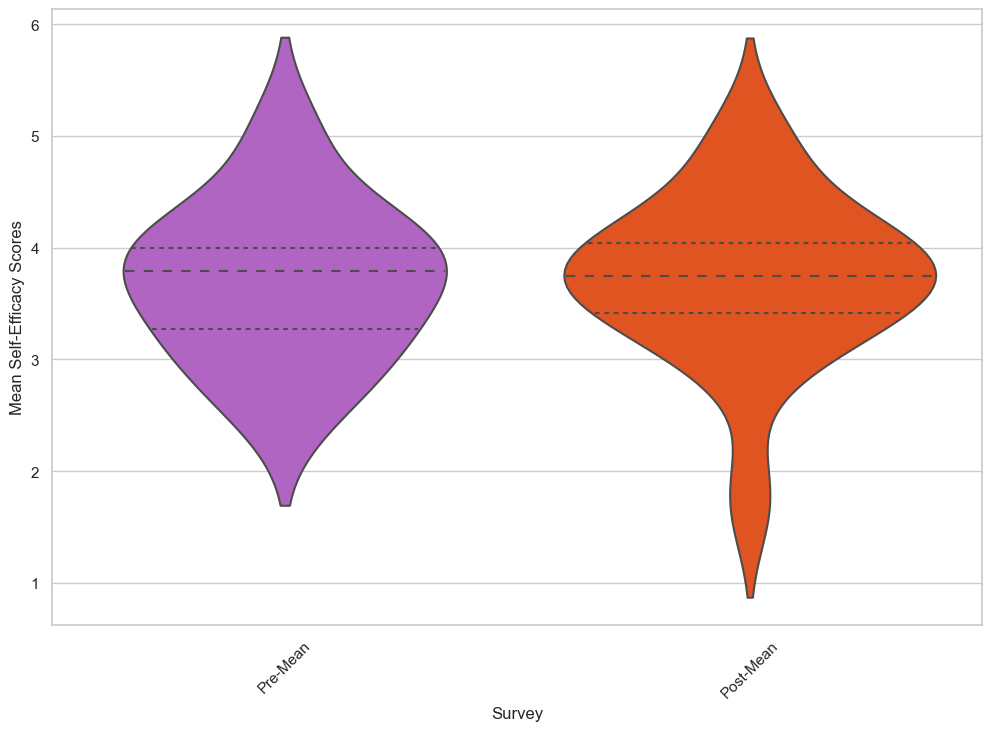}
        \caption{Distribution of Pre-study and Post-study self-Efficacy scores}
        \label{fig:violin}
    \end{subfigure}
    \hfill
    \begin{subfigure}[b]{0.45\textwidth}
        \centering
        \includegraphics[width=\textwidth]{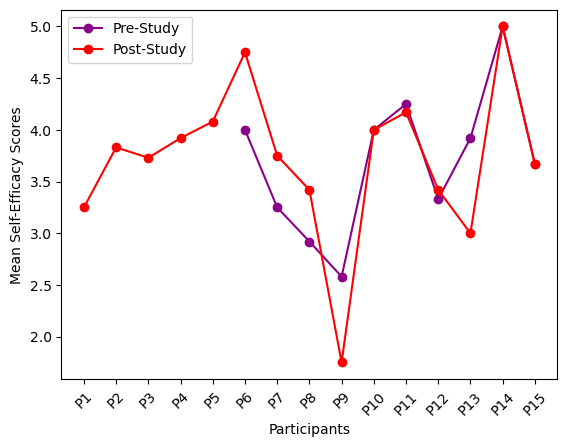}
        \caption{Pre- vs. Post-study self-efficacy scores by participant}
        \label{fig:line}
    \end{subfigure}
    \caption{Self-efficacy of participants}
    \label{fig:both_plots}
\end{figure}

Figure \ref{fig:both_plots} presents the participants' perception of self-efficacy before and after using \genai. Figure \ref{fig:violin} shows the distribution of perception of self-efficacy before and after the study. Given the small number of observations, we cannot draw any statistically-sound conclusion about any difference between the distributions.

Figure~\ref{fig:line} compares the self-efficacy of the participants before and after the experiment. 
It shows that P6 experienced a noticeable increase in self-efficacy, from a self-efficacy score of $4.00$ in pre-study to $4.75$ after the study, indicating a potential positive impact of using \genai on self-efficacy. 
P6 completed all three programming tasks successfully and demonstrated a mix-model behavior by independently addressing two tasks and using a hybrid approach with \genai for one task. The increase in self-efficacy scores suggests that employing \genai was effective in improving her perception of self-efficacy. 

In contrast, self-efficacy in P9 decreased from $2.58$, before the study, to $1.75$ after the study. 
Although P9 completed all programming tasks successfully, she copied the full description to the plug-in, asked ChatGPT for the answer, and pasted the results as the submission. 
Similarly, this occurred for P13, who began with a pre-study self-efficacy score of $3.92$, however, the post-study mean score dropped to 3.00. 
He submitted an incorrect solution for one task and did not finish the other two tasks, she used \genai in a step-by-step approach.

\section{Threats to Validity}
\textbf{Replicability}
\emph{Can others replicate our results?}
Because \genai and its usage and acceptance are rapidly emerging, we do not know if the results of this study will be replicable due to changing student behaviors and rapid advances in \genai. We encourage other researchers to replicate our study for different populations to check for emerging behaviors of students when using \genai. 

\noindent{\textbf{Internal}}
\emph{Did we skew the accuracy of our results with how we collected and analyzed data?}
We used open coding to analyze the data and two authors reviewed all the data to make sure that the coding of videos adhered to the codebook that was created. 

\noindent{\textbf{External}}
\emph{Do our results generalize?}
Because our study has a small sample size (n=15) relative to the overall student population, it is not possible to generalize to all students' behavior. 
However, we attempted to have diversity based on gender.
The bias of drawing conclusions from self-selection bias remains a threat to validity.

\section{Discussion and Concluding Remarks}

In this section, we discuss the potential implications of our findings for computing education research and practice.  

\subsection{Interactions with \genai}
In this study, our objective was to investigate how CS1 students utilize \genai for programming questions to assess whether \genai serves as a help-seeking tool for beginners in programming. As observed in other recent studies ~\cite{kazemitabaar2023novices, prather2024widening}, our findings indicate that a large number of \genai users rely on providing full descriptions of programming questions to find solutions without making sufficient effort on their own, even under supervision. This trend aligns with the patterns of over-reliance on LLMs identified by Kazemitabaar \etal ~\cite{kazemitabaar2023novices}, particularly among novices using the 'AI single prompt' approach, which resulted in lower performance on subsequent tasks.
The observed behavior raises concerns about the potential overreliance on \genai in educational settings, where students might increasingly rely on \genai to provide all solutions to the detriment of their learning. This echoes the observations of Fernandez and Cornell ~\cite{fernandez2024cs1}, who emphasized the need for careful integration of AI-driven code generation tools to avoid such overreliance.
Help-seeking is crucial for students to grasp new concepts, acquire skills, and tackle challenges in their computing courses \cite{hou2024effects}. However, when participants use full descriptions of programming tasks as prompts and accept complete \genai-generated responses, \genai may not effectively fulfill its role as a help-seeking tool that constructively aids struggling students, but an oracle that does the learners' job for them. This concern was also raised by Jo\v{s}t \etal \cite{jost2024impact}, highlighting the need for instructional strategies that emphasize breaking down problems and leveraging \genai for incremental learning.
Our findings, alongside those of Prather \etal ~\cite{prather2024widening}, suggest that some students may struggle with new metacognitive difficulties when using generative AI tools, including being conceptually behind in course material but unaware of it due to a false sense of confidence. Further exploration is needed to understand the underlying reasons for this undesired behavior and to encourage a more constructive use of \genai that promotes deep understanding and problem-solving skills in computing education.

Our observations reveal two main types of behavior in problem solving: iterative and linear.  Students who employed an iterative approach refined their prompts to achieve correct answers, while students who used a linear approach used the full description of the problems to find answers directly. This behavioral split reflects the findings of Vadaparty \etal~\cite{vadaparty2024cs1llm}, who noted similar patterns in student interactions with \genai in a CS1 course. The iterative approach can improve learning by encouraging deeper engagement with problem-solving processes, whereas the linear approach may indicate a tendency to seek quick fixes.

Some prompts were related to coding concepts, indicating that novices struggle the most to understand programming concepts. This is consistent with the findings of Liu \etal~\cite{liu2024teaching}, who found that students often used \genai tools to clarify the concepts of coding and the logic of the program. Our study further suggests that while \genai can assist with the coding concepts, additional instructional support is needed for the logic and debugging of the program.

Our results show that in 30\% of prompts, users accepted the 
\genai response, while in 70\%, AI responses were used to learn concepts. 
This dual role of \genai as both an answer provider and a personal tutor aligns with the observations by Prather \etal~\cite{prather2024widening} regarding the mixed impact of \genai on student learning, where \genai facilitated both understanding and dependency. Effective \genai integration should balance assistance with promoting independent problem-solving skills.


We can categorize the nature of interactions with \genai based on the reliance of a prompt on prior prompts into two groups: (1) stateless and (2) stateful. 
In the stateless interactions, individual prompts are independent of the prior prompts, hence prompts can be interpreted and answered independently.
However, in stateful interactions, the prompt assumes that \genai uses the history of the user's interaction.   
For instance, P7 used stateful interaction, where he asked ``So, what was wrong with my initial code?''.

We observed that seven participants P4, P7, P9, P10, P13, P14, and P15 employed the same approach on all tasks. We call their strategy for problem-solving, \textit{single-approach} strategy. In contrast, P1, P2, P3, P5, P6, P8, P11, and P12 used different approaches for different tasks that we call \textit{mixed-approach}. 
As generative AI becomes more commonplace and more learners can access them for diverse sets of topics, a more granular investigation of these approaches becomes more important to guide pedagogy.

\subsection{Self-efficacy and \genai}

Our observations suggest that users who used full descriptions of the questions may prefer smooth interactions without challenges. This behavior correlates with lower self-efficacy scores, similar to the findings of Xue \etal~\cite{xue2024chatgpt} and Prather \etal~\cite{prather2024widening}, who observed that students with lower self-efficacy tend to rely more heavily on \genai tools. 
Overall, some users' self-efficacy levels increased after programming with \genai. This suggests the varied impacts of integrating \genai in educational contexts, shaping self-efficacy outcomes according to individual learning strategies and initial confidence levels. Enhancing self-efficacy through scaffolded \genai interactions could support more confident and independent problem-solving.

\subsection{Implications for Computing Education}
The current advancement in prompt engineering focuses on productivity, aiming to help developers find final solutions quickly. 
However, in an educational context, the goal is not merely to reach a solution, but to ensure that students achieve the learning objectives. 
This goal is inherently different from prompt engineering for productivity improvement, which aims to minimize and eradicate failures, a proper learning strategy may require expecting or even encouraging a significant amount of purposeful failures along the way.
Thus, prompt engineering research prioritizing individual learning, as seen in works like Jin \etal~\cite{jin2024teach}, should be further investigated.

Our results suggest that, even in the physical presence of a researcher and with knowledge of being recorded, in a considerable number of cases, the participants directly resorted to \genai to provide solutions to the tasks without making any attempts on their own. 
This undesirable way of help-seeking, if used as the default approach in solving homework problems, can negatively impact students' learning.  
This suggests the need for students to improve their self-regulation skills, e.g. self-monitoring~\cite{schunk2012self} to monitor and reflect the intensity and frequency of their use of \genai. Similarly, \genai educational tool builders should consider supporting such strategies that enhance students' self-regulation and help-seeking behavior when they use \genai.

The findings of Margulieux \etal~\cite{margulieux2024self} are particularly relevant here. In their study, they found that some students used \genai to support and not replace their critical thinking and problem solving. However, they also noticed that the weakest students tended to use \genai earlier in the problem solving process. Their findings mirror those in our study. However, Margulieux \etal were somewhat optimistic that their findings meant that users who needed help and support could receive it. Our findings, as well as those by Prather \etal~\cite{prather2024widening}, show that lower-performing students using \genai earlier in the problem solving process is likely an indication of overreliance and a failure to properly self-regulate with these tools.

One can imagine that a possible remedy for this problem is to devise novel homework assignments that can be difficult for \genai to solve. 
However, as \genai tools become more sophisticated and powerful, the arms race between educators and \genai seems to be a losing proposition, especially in introductory courses such as CS1 that only include basic algorithmic thinking that there is an upper limit to the appropriate complexity of problems ~\cite{prather2023robots, denny2024cacm}. 
To discourage students from using prompts to find complete solutions to homework problems, we should investigate novel \emph{types} of problems that are compatible with the era of \genai. 
Recent work such as Prompt Problems~\cite{denny2023promptly} that uses the graphical representation for problem description instead of a textual description is a step in this direction. 
Even though the newest models with visual modality have been able to solve Prompt Problems \cite{hou2024morerobots}, they remain useful as a way to scaffold students usage of \genai by helping them learn problem decomposition, iterative problem solving, describing a problem, and prompt engineering \cite{prather2024interactions}.

An intriguing observation from our study was that most participants did not frequently execute their programs during development to verify the correctness or identify syntax errors. 
This lack of regular testing suggests a gap in their understanding of the importance of iterative debugging in the programming process. 
Consequently, students may miss out on early detection of mistakes, which can lead to more complex issues and frustration later in the development cycle. 
Addressing this behavior through new pedagogical tools such as \cite{ferdowsi2024validating:chi:leap} that visualizes the value of variables as students develop their solutions could improve students' coding practices and overall comprehension of programming concepts.

\section*{Acknowledgments}
This material is based upon work supported by the U.S. National Science Foundation under Grant No. 2225373. Any opinions, findings, and conclusions or recommendations expressed in this material are those of the author(s) and do not necessarily reflect the views of the National Science Foundation.

\bibliographystyle{ACM-Reference-Format}

\bibliography{bib} 
\end{document}